\documentclass[12pt]{iopart}
\usepackage{graphicx}
\usepackage{hyperref}
\hypersetup{
    unicode=false,          
    pdftoolbar=true,        
    pdfmenubar=true,        
    pdffitwindow=false,     
    pdfstartview={FitH},    
    pdftitle={GWs from CCSN: CHIMERA Simulations},    
    pdfauthor={Konstantin N Yakunin},     
    pdfsubject={Subject},   
    pdfcreator={Creator},   
    pdfproducer={Producer}, 
    pdfkeywords={supernovae: general - numerical simulations - hydrodynamics - 
                          equation of state - neutrinos - gravitational waves}, 
    pdfnewwindow=true,      
    colorlinks=true,       
    linkcolor=red,          
    citecolor=blue,        
    filecolor=magenta,      
    urlcolor=cyan           
}

\newcommand{\ApJ}{{\it Astrophys. J. }}
\newcommand{\AAJ}{{\it Astron. Astrophys. }}

\begin{document}

\title[Gravitational Waves from Core Collapse Supernovae]
       {Gravitational Waves from Core Collapse Supernovae}

\author{Konstantin N Yakunin$^1$, 
Pedro Marronetti$^1$, 
Anthony Mezzacappa$^{2,3}$, 
Stephen W Bruenn$^1$, 
Ching-Tsai Lee$^3$, 
Merek A Chertkow$^3$, 
W Raphael Hix$^{2,3}$, 
John M Blondin$^4$, 
Eric J Lentz$^{2,3}$, 
O E Bronson Messer$^5$, 
and Shin'ichirou Yoshida$^6$}

\address{$^1$ Physics Department, Florida Atlantic University, Boca Raton, FL 33431-0991}
\address{$^2$ Physics Division, Oak Ridge National Laboratory, Oak Ridge, TN 37831-6354}
\address{$^3$ Department of Physics and Astronomy, University of Tennessee, Knoxville, TN 37996-1200} 
\address{$^4$ Department of Physics, North Carolina State University, Raleigh, NC 27695-8202}
\address{$^5$ National Center for Computational Sciences, Oak Ridge National Laboratory, Oak Ridge, TN 37831-6354}
\address{$^6$ Department of Earth Science and Astronomy, University of Tokyo}

\ead{cyakunin@fau.edu, pmarrone@fau.edu} 

\begin{abstract}
We present the gravitational wave signatures for a suite of axisymmetric core collapse supernova models with progenitors masses
between 12 and 25 M$_\odot$. These models are distinguished by the fact they explode and contain essential physics 
(in particular, multi-frequency neutrino transport and general relativity) needed for a more realistic description. 
Thus, we are able to compute complete waveforms (i.e., through explosion) based on non-parameterized, first-principles models. 
This is essential if the waveform amplitudes and time scales are to be computed more precisely. 
Fourier decomposition shows that the gravitational wave signals we predict should be
observable by AdvLIGO across the range of progenitors considered here.
The fundamental limitation of these models is in their imposition of axisymmetry.
Further progress will require counterpart three-dimensional models.
\end{abstract}



\section{Introduction}

Core collapse supernovae are among the sources that will produce gravitational waves (GWs) detectable by GW observatories 
around the globe. In particular, a Galactic supernova event will likely produce a signal well within Advanced LIGO's \cite{AdvLIGO} bandpass across a broad range of frequencies.
Gravitational waves from supernovae will arise from a variety of phenomena given their multidimensional, multi-physics character. 
These phenomena include fluid instabilities in the proto-neutron star, neutrino-driven convection beneath the supernova shock wave, 
the standing accretion shock instability (SASI), deceleration at an aspherical shock, and aspherical neutrino emission. 
Obviously, two-, and ultimately three-, dimensional models will be required to capture the GW emission from such phenomena. 
Moreover, the explosion dynamics, and ultimately the computation of the GW emissions as well, 
will require sufficient realism in the treatment of core collapse supernova multi-physics. For a comprehensive survey of the field see the recent review by Ott {\cite{Ott2009}} and references therein.

\section{The CHIMERA Code}

Our CHIMERA code has five major components: hydrodynamics, neutrino transport, self-gravity, a nuclear equation of state, 
and a nuclear reaction network (see \cite{Bruenn09} for details). The hydrodynamics 
is evolved via a Godunov finite-volume scheme---specifically, a Lagrangian remap 
implementation of the Piecewise Parabolic Method (PPM) \cite{Colella84}. 
Neutrino transport along our radial rays is computed by 
means of multigroup flux-limited diffusion (in the ``ray-by-ray-plus`` approximation \cite{Buras06}), with a flux limiter that has been 
tuned to reproduce Boltzmann transport results to within a few percent \cite{Liebendorfer04}.
A spectral Poisson solver is used to determine 
the gravitational field \cite{Muller95}, with general relativistic (GR)
corrections to the spherical component \cite{Marek06}. Details of the integration of this gravitational 
framework with the hydrodynamics will be given in \cite{Bruenn10}.
The equation of state (EOS) of Lattimer \& Swesty \cite{Latt91} 
is currently employed for matter in NSE above $1.7\times10^8 ~{\rm g/cm}^3$. Below this 
density, matter in NSE is described by 4 species (neutrons, protons, helium, 
and a representative heavy nucleus) 
in a corrected and improved version of the Cooperstein EOS \cite{Cooperstein85}, 
extended to regions where the composition is determined externally by a reaction network.  
Continuity with the LS-EoS is achieved by use of a common electron-positrion EOS (a revised and extended version of that in \cite{Cooperstein85}) 
and establishment of a common zero for the mass energy.
For regions not in NSE, an EOS with a nuclear 
component consisting of 14 $\alpha$-particle nuclei from $^4$He to $^{60}$Zn, protons, neutrons, and an iron-like nucleus is used. 
An electron--positron EOS with 
arbitrary degeneracy and degree of relativity spans the entire density--temperature 
regime of interest. The nuclear composition in the non-NSE regions of these 
models is evolved by the thermonuclear reaction network of Hix \& Thielemann \cite{Hix99}. 
\begin{figure}
\begin{center}
\includegraphics[scale=0.8]{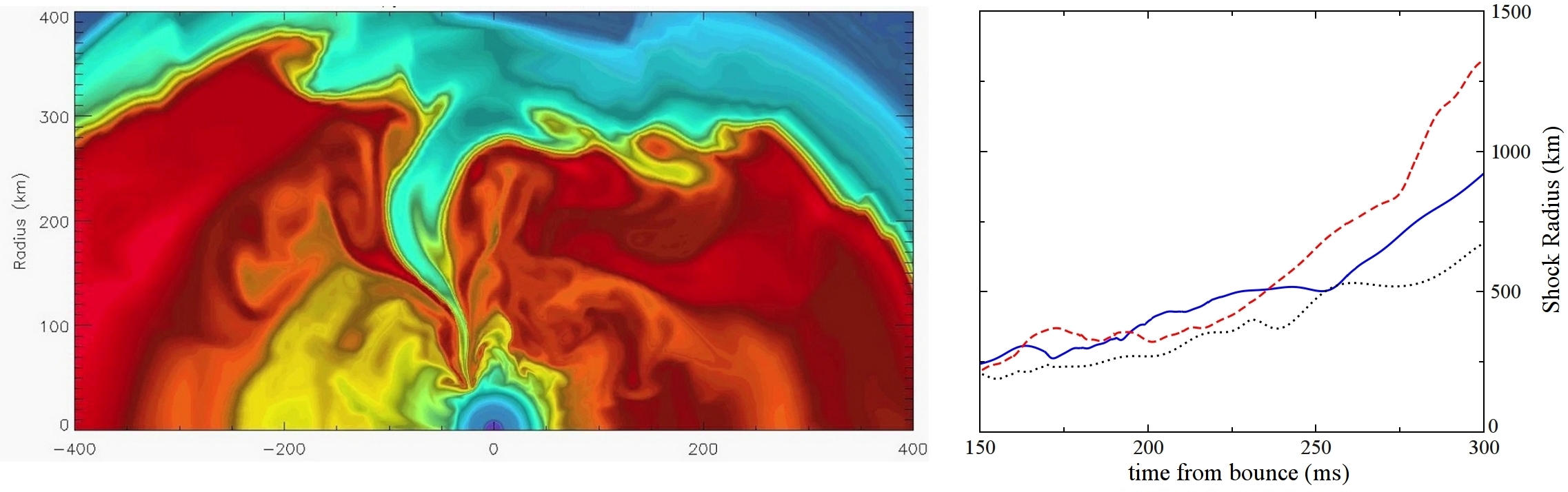}
\caption{{\it Left:} Entropy distribution at 244 ms after bounce for the 15 M$_{\odot}$ model. 
A large, low-entropy (blue-green) accretion funnel at an angle quasi-orthognal to the symmetry axis and  high-entropy (yellow-orange-red) 
outflows below the shock, along the symmetry axis, are evident.
{\it Right:} Shock radius as a function of time for three regions: the north pole (solid blue), the equatorial plane (dotted black),
 and the south pole (dashed red).}
\label{snapshot}
\end{center}
\end{figure}
While Eulerian schemes are preferred for regions with violent turbulence, they have a disadvantage: 
the history of field variables for a given parcel of material, crucial for nucleosynthesis, is lost. 
To compensate for this loss, and to allow post-processed nuclear network computations, 
the tracer (or test) particle method \cite{Lee08} has been implemented in CHIMERA.
The tracer particles are equally distributed on the spherical grid (40 particles/row x 125 rows) 
at the pre-collapse phase and follow the flow in the course of the Eulerian simulation, 
recording their temperature and density history by interpolating the corresponding quantities from the underlying Eulerian grid \cite{Lee08}. 
Each particle is assigned a constant mass ($1/5000$ of the progenitor mass) 
and the GW signal it produces is calculated taking the quadrupole integral.
Comparing the GW corresponding to a given group of tracers with the signal produced by the bulk matter motion allows us 
to identify what part of the fluid generates a specific GW feature.

\section{Gravitational Wave Extraction}

Stochastic matter motion and anisotropic neutrino emission during the explosion generate GWs.
The transverse-tracefree part (TT) of the gravitational strain is written as
\begin{eqnarray}
 h_{ij}^{TT} = \frac{1}{r}\sum_{m=-2}^{m=2}
  \left(\frac{d}{dt}\right)^2I_{2m}\left(t-\frac{r}{c}\right)
~ f^{2m}_{ij}, \nonumber
\end{eqnarray}
where the mass quadrupole (as a function of retarded time) is computed by
\begin{eqnarray}
 I_{2m} = \frac{16\pi G}{5c^4}\sqrt{3}\int \tau_{00} Y^*_{2m}r^2dV, 
\nonumber
\end{eqnarray}
with $\tau_{00}$ the corresponding component of the linearized stress-energy tensor, 
and $f^{2m}(\theta, \phi) $ the spherical harmonics. 
In the weak-field case, we approximate $\tau_{00} \simeq \rho$, where
$\rho$ is the rest-mass density. Following the Finn--Evans approach \cite{Finn90} to reduce 
the second time derivative $A_{2m} \equiv \frac{d^2}{dt^2}I_{2m}= \frac{d}{dt}N_{2m}$ 
and using the continuity equation (Blanchet \etal \cite{Blanchet90}), we can calculate $N_{2m}$ as in Equation (34) in 
 \cite{Finn90}.
In axisymmetric cases, $N_{20}$ is the only non-null component, and we evaluate its time derivative numerically.
The wave amplitude is related to the dimensionless gravitational strain,
$h_+$, by
\begin{equation*}
   h_+=\frac{1}{8}\sqrt{\frac{15}{\pi}}\sin^2\theta\frac{A_{20}}{r},
\end{equation*}
where $r$ is the distance to the source and $\theta$ is the angle between the symmetry axis and the observer's line of sight 
(we will assume $\sin^2\theta = 1$).
To compute the GWs produced by anisotropic axisymmetric neutrino emission, we use Epstein's \cite{Epstein78} and 
M{\"u}ller \& Janka's formalism \cite{Muller97}:
\begin{equation*}
 h_{\nu}^{TT} = \frac{4G}{c^4 r}\int_{0}^{t} dt'
  \int_{0}^{\pi} d\theta'~\Psi(\theta')
  \frac{dL(\theta',t')}{d\Omega'},
\label{hnu}
\end{equation*}
with $\Psi(\theta)$ given in \cite{Kotake07}. The direction-dependent  
differential neutrino luminosity, $dL/d\Omega$, is calculated at the outermost radial grid zone. 
In order to determine the detectability of the GWs, 
we calculate the characteristic GW strain for a given frequency $f$ \cite{Flanagan98} using
\begin{eqnarray}
  h_c\left( f \right)= \frac{1}{r}\sqrt{\frac{2}{\pi^2}\,\frac{G}{c^3}\,\frac{dE_{GW}\left(f\right)}{df}} & 
  ~~\textrm{and}~~  
   \frac{dE_{GW}\left(f\right)}{df}=
     \frac{c^3}{G}\,\frac{\left(2\pi f\right)^2}{16\pi} \left|\tilde{A}_{20}\left(f\right)\right|^2, \nonumber
\end{eqnarray}
where $dE_{GW}\left(f\right)/df$ is the GW energy spectrum and
$\tilde{A}_{20}\left(f\right)$ is the Fourier transform of $A_{20}\left(t\right)$.

\begin{figure}
\begin{center}
\includegraphics[scale=1.0]{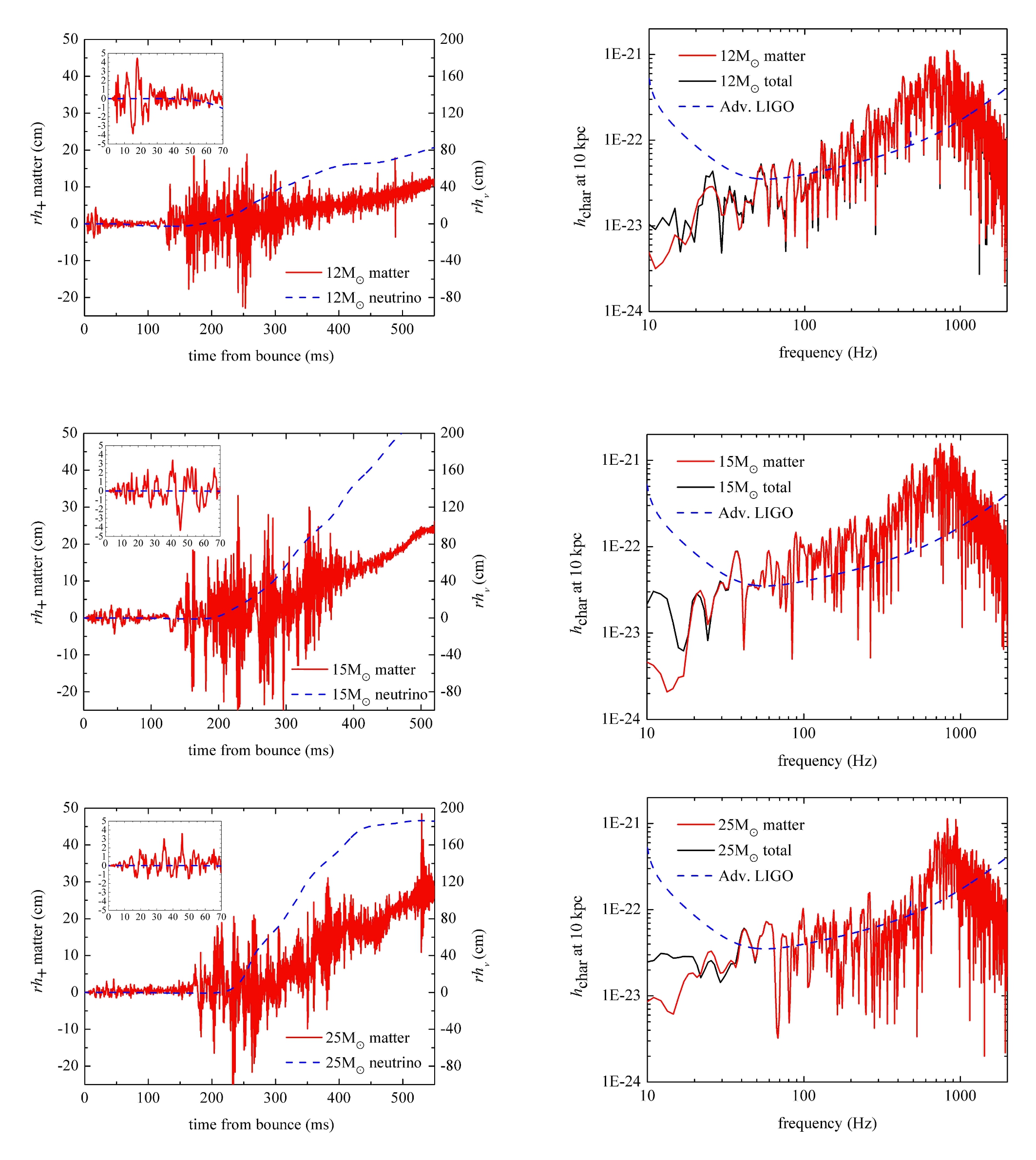}
\caption{ The left column shows the GW strain times the distance to the observer vs. post-bounce time for non-rotating progenitors
 of 12, 15, and 25 $M_{\odot}$. 
The signal is split into matter- (red-solid) and neutrino-generated (blue-dashed) signals. 
Note that the scales are different for these two signals.
The insets show the first 70 ms after bounce. 
The right column shows the corresponding characteristic strain for both the matter (red) and the total (black) signals, 
compared to the AdvLIGO sensitivity curve.}
        \label{GW_4_models}
\end{center}
\end{figure}

\section{Gravitational Waveforms}

We performed axisymmetric two-dimensional simulations beginning with 12, 15, and 25 $M_\odot$ 
non-rotating progenitors \cite{Woosley07} and resolutions of 256 (adaptive) radial and 256 angular zones. 
The radial grid ranges from $0$ km to $1.88\cdot 10^4$ km. 
The left panel of Figure \ref{snapshot} shows an entropy snapshot of the 15 $M_\odot$ model.
Successful explosions are obtained in all cases, with the longest running model (the 25 M$_\odot$ model) 
having an explosion energy of 0.7 B (and still growing) 1.2 seconds after bounce.
Details are provided in \cite{Messer08,Bruenn09}.

While the GW emissions we predict differ in detail from model to model, a clear GW signature, composed of four parts
 (left column of Figure \ref{GW_4_models}) emerges: 
1) A {\bf prompt signal}: an initial and relatively weak signal that starts at bounce and ends at between 50 and 75 ms post-bounce.
 2) A {\bf quiescent stage} that immediately follows the prompt signal and ends somewhere between 125 ms and 175 ms after bounce.
 3) A {\bf strong signal}, which follows the quiescent stage and is the most energetic part of the GW signal.
 This stage ends somewhere between 350 ms and 450 ms after bounce. 
4) A {\bf tail}, which starts before the end of the strong signal at about 300 ms 
after bounce and consists of a slow increase in $rh_{+}$. This tail continues to rise at the end of our runs.

Waveforms covering the first three of four phases (prior to explosion) have been computed by Marek \etal \cite{Marek09},
 and waveforms covering all four phases and based on parameterized explosions were reported in the work of Murphy \etal \cite{Murphy09}.
 The overall qualitative character of the GW signatures shown in \cite{Murphy09} reflect what is shown in Figure \ref{GW_4_models}. 
The work presented here takes the natural next step beyond this earlier, foundational work.
 A more precise prediction of the GW amplitudes and the timescales associated with each of the four phases requires 
a non-parameterized approach. Even in the case of a non-parameterized approach, prior to evidence of an explosion 
it is difficult to assess whether or not the amplitudes and timescales are well determined. 
Thus, the non-parameterized {\it explosion} models studied here enable us to predict all four phases of the GW emission 
and their amplitudes and timescales with some confidence.
\begin{figure}
\begin{center}
\includegraphics[scale=0.7]{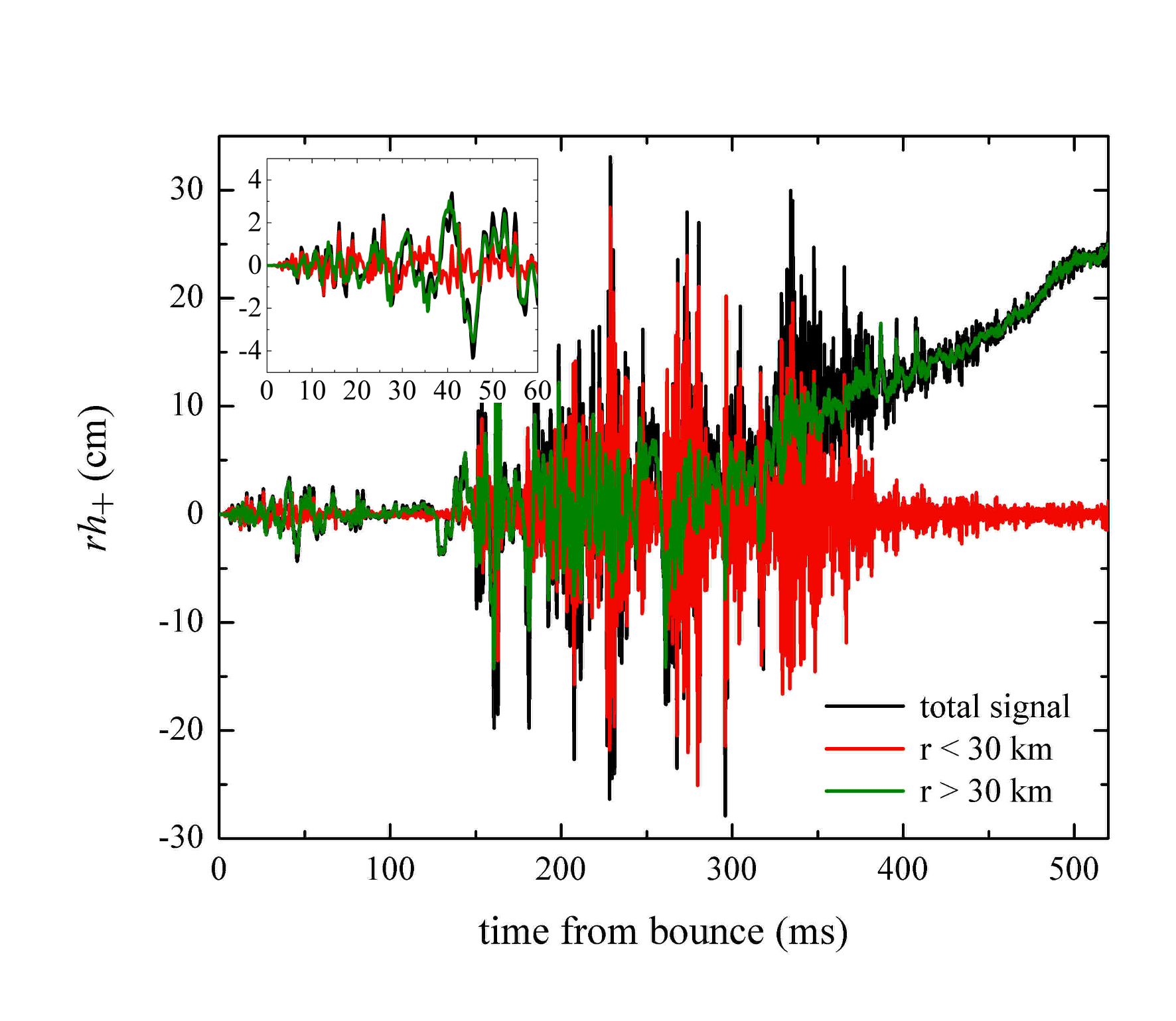}
\caption{Contributions to the matter-generated GW signal from two different regions for the 15 $M_\odot$ model: the PNS ($r< 30$ km) and
the region above the PNS ($r > 30 {\rm km}$). 
The latter includes the region of neutrino-driven convection, the SASI, and the shock.}
        \label{GW_s15_zones}
\end{center}
\end{figure}

\begin{figure}
\begin{center}
\includegraphics[scale = 1]{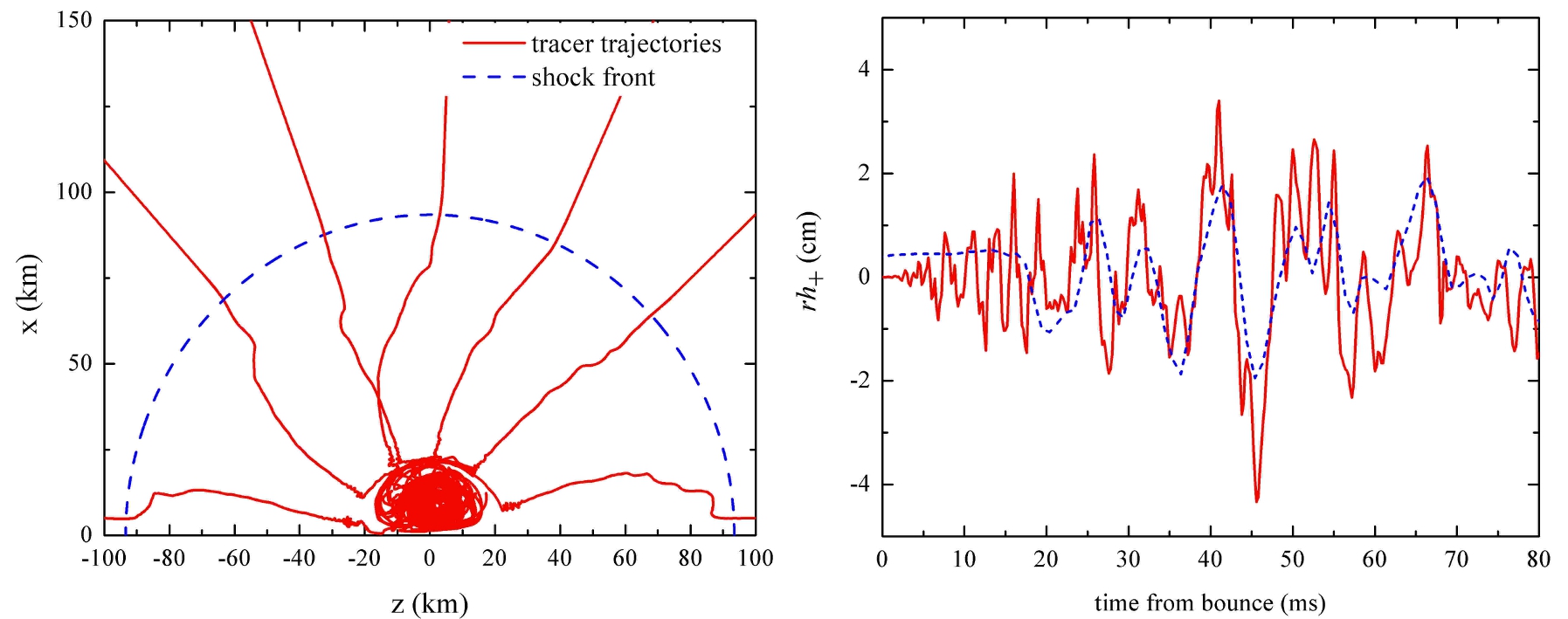} 
\caption{{\it Left:} Trajectories of the tracer particles. 
          It is shown the clear deflections of infalling particles through the shock that collectively produce 
          low-frequency high-amplitude component of the GW signal shown on the right panel.  
          {\it Right:} Comparison between the matter signal (solid red) and signal calculated using the tracers (dashed blue). 
          Both panels correspond to our 15 $M_\odot$ simulation.}
        \label{Tracer}
\end{center}
\end{figure}

The prompt signal is generated by {\it two} independent phenomena: prompt convection inside the PNS generates 
a high-frequency signal that is superimposed on a lower-frequency component, seen in the insets of Figure \ref{GW_4_models}. 
There is a hint of this in the inset of Figure \ref{GW_s15_zones}, where the signal for our 15 $M_\odot$ run has been 
split into the contributions from two different regions, but it is in the tracer analysis of Figure \ref{Tracer} (right) where this becomes evident. 
The matter-generated GW (solid red) is closely tracked by the GW generated by the infalling tracer particles deflected by the shock (dashed blue),
some of which are shown in the left panel.
The low-frequency signal from ~20 ms to 60 ms after bounce originates at the shock 
radius, which is at $\sim$100 km at this time and well outside the PNS. 
In the past, authors attributed the prompt signal to convection only \cite{Marek09, Murphy09}.

The quiescent stage corresponds to the period after prompt convection has ceased and prior to the development of neutrino-driven 
convection and the SASI. It is followed by a strong signal produced by the development of both. The strong signal is dominated by 
SASI-induced funnels impinging on the PNS surface. 
It shows evidence of two components (also described in \cite{Marek09, Murphy09}): The low-frequency component arises from the modulations in the shock radius as the SASI develops and evolves. The right panel of Fig. \ref{snapshot} shows the first cycle of this modulation at 175ms with the maximum (minimum) of the north (south) pole radius and at about 210ms with the reverse situation. The high-frequency component is generated when the SASI-induced accretion flows strike the PNS (Fig. \ref{snapshot} left). The shock modulations affect the kinetic energy of the accretion flows and, consequently, the amplitude of the GWs generated when these flows hit the PNS.
Hence the high-frequency modulations are beneath a low-frequency envelope. 

All of our GW signals end with a slowly increasing tail, which reflects the gravitational memory associated with 
accelerations at the prolate outgoing shock (see also Fig. 5 in \cite{Murphy09}).
The tail continues to rise at the end of our runs because the explosions are still developing and strengthening. 
The 15 M$_\odot$ model GW is shown in Figure \ref{GW_s15_zones}, 
where the explosion starts at $\sim 300-350$ ms after bounce, and by 400 ms the signal from the PNS has largely ceased.

Focusing now on $h_{\rm char}$ (Figure \ref{GW_4_models} right), 
it is important to note that the peak at $\sim 700 - 800$ Hz is associated with the high-frequency component of $rh_+$,
 which in turn is associated with the downflows hitting the PNS surface, as discussed above. 
A precise association of the signal at lower frequencies with phenomena in the post-bounce dynamics will require
 a detailed analysis using tracer particles and will be left to a subsequent paper \cite{Bruenn10}.
 The lower-frequency modulations (the envelope) in $rh_+$, which in turn are associated with the SASI-induced 
shock modulations will certainly be an important component of this lower frequency signal. 
Finally, we note that the possible dependence of the AdvLIGO-observable signal between 100 Hz and 700 Hz 
on the progenitor mass also requires further investigation and a detailed discussion, which will be presented in \cite{Bruenn10}.

The amplitudes of the GWs from neutrino emission are negative from bounce to $\sim 180-220$ ms after bounce and then increase 
dramatically, becoming positive throughout the end of the simulation. The positive sign is consistent with 
a relative dominance of neutrino emission along the polar over the equatorial regions \cite{Kotake07}. 
The change in sign from negative to positive
correlates with the formation of the funnel-like downflows of dense matter, which increase neutrino opacities 
in the equatorial plane (orthogonal to the symmetry axis; see Figure \ref{snapshot}). 
Note that the amplitude of the neutrino-generated GW signal is much larger than the matter-generated GW signal.
 However, these GWs have relatively low frequencies, and their contribution to the total characteristic strain is 
only significant at frequencies below $20$ Hz (see also \cite{Muller97, Kotake07, Burrows96, Kotake09}).

\begin{figure}
\begin{center}
\includegraphics[scale=1.0]{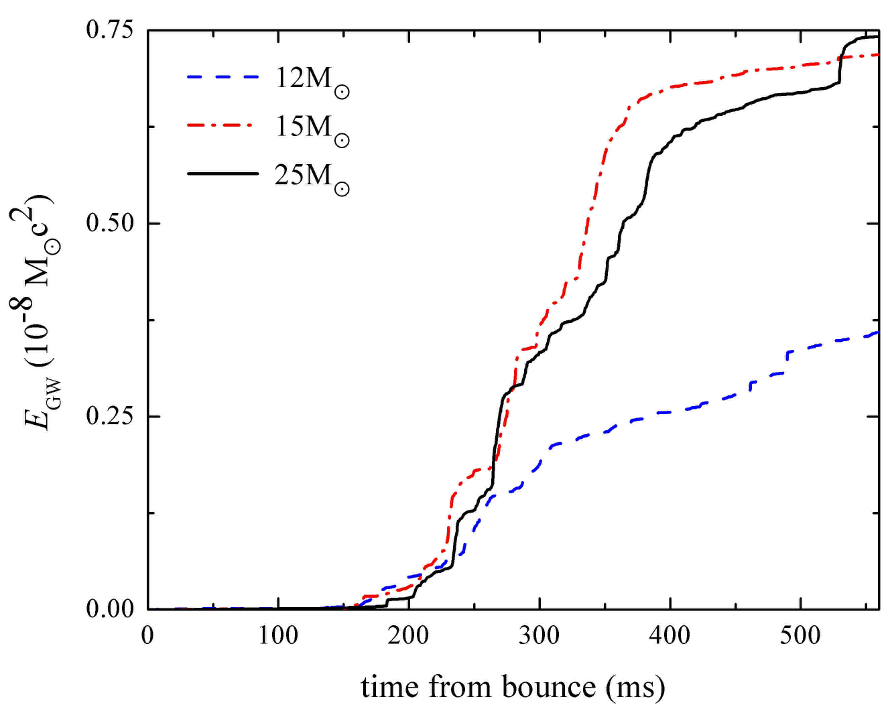} 
\caption{Energy emitted by GWs during the first 500 ms after bounce for all three models presented here.}
        \label{Egw}
\end{center}
\end{figure}

Our GW predictions for the 15 M$_\odot$ case can be compared to those of Marek \etal \cite{Marek09}
 given that both groups implement similar
treatments of the neutrino transport and GR corrections to the gravitational field, and include essentially the same 
overall multi-physics in their models. The two groups are in agreement with regard to the time scales of the different 
(pre-explosion) GW phases, the amplitude of the prompt signal, and the peak in the GW spectrum at $\sim 700 - 800$ Hz. 
They differ, however, in their predictions for the amplitude of the strong signal, where our results are about twice as large. 
These differences will be investigated. They likely arise in part due to the different progenitors used, which in turn alters the 
time scale to explosion and, consequently, the amplitude of the GW signal in the strong-signal phase at any instant of time.

A comparison with Murphy \etal \cite{Murphy09} is more difficult given that their models are parameterized 
(i.e., their neutrino luminosities are kept isotropic and constant throughout their simulations). 
As a result, we can expect the GW amplitudes, time scales, and 
frequency peaks to be different. Murphy \etal\cite{Murphy09} present a simple model for the GW strain during the 
strong-signal phase, where the GW amplitude, caused by the downdrafts hitting the PNS surface, is proportional to the 
downdraft frequency $f_p$ and velocity $v_p$, which, in turn, depend on the compactness of the PNS. In light of this 
model, the larger amplitudes and characteristic frequencies we see might be attributed in part to our use of an effective GR 
potential and the soft Lattimer \& Swesty EOS
rather than a Newtonian potential and the stiff Shen EOS \cite{Shen98}.

The total emitted GW energy is shown in Figure \ref{Egw}. For the more massive progenitors, 
all of the GW energy is emitted between 200 ms and 400 ms after bounce. 
For the 12 M$_\odot$ case, the GW energy is emitted more slowly, consistent with the fact 
that the explosion in this case unfolds more slowly \cite{Bruenn09}.
 Our predictions are 20 to 50 times larger than those of Murphy et al. \cite{Murphy09},
 but this is consistent with our waveforms having two to three times the amplitude and a higher frequency than 
the signals they predict.

\section{Summary and Conclusions}

We present gravitational waveforms computed 
in the context of 2D core collapse supernova simulations performed with 
the CHIMERA code for non-rotating 12, 15, and 25 $M_\odot$ progenitors. 
We calculate the contribution to the signals produced by both baryonic matter motion
 and anisotropic neutrino emission up to 530 ms after bounce for all three progenitors. 
Given the development of non-parameterized explosions in our models, we are able to compute the waveforms 
through explosion and to determine more precisely the pre-explosion amplitudes and timescales. 
 Given our use of tracer particles, we are able to decompose the GW signatures and determine 
 which phenomena contribute to specific components of the waveforms. 
This allowed us to identify an additional source for the prompt signal (in the past solely attributed to prompt convection) the deflection of infalling matter through the shock.

Our waveforms exhibit a characteristic signature. Namely, the signal develops in four stages. 
There is 1) a relatively short and weak {\bf prompt signal}, 2) a {\bf quiescent stage}, 3) a {\bf strong signal} 
where most of the GW energy is emitted and, 4) a slowly increasing {\bf tail}. 
We predict signatures with sufficient strength to be readily observable by Advanced LIGO for a Galactic event, 
and the peak in the observable spectrum stems from the accretion downflows driven by the SASI.

The results presented here are preliminary: a new set of 2D simulations performed 
with an enhanced version of our CHIMERA code is currently ongoing \cite{Bruenn10}.
However, while a number of approximations are made in the CHIMERA code, 
the more important limitation in these models is their restriction to axisymmetry. Three-dimensional models are required. 
We anticipate that the greatest change to our gravitational waveform predictions in moving to 3D will be in the phase-4 tail. 
Prolate explosions are often seen in axisymmetric simulations, where artificial boundary conditions must be imposed 
that prevent the turnover of material along the symmetry axis. With axisymmetry removed, we expect a significant change 
in the evolution of the explosion tail: in its magnitude, and perhaps even in its sign. 
And no doubt there will be quantitative changes to the amplitudes and timescales associated with earlier phases, 
particularly in the strong signal arising from the SASI motions. 
In 3D, the SASI will likely be dominated by spiraling flows \cite{Blondin07}, fundamentally different than 
the sloshing modes that dominate in the axisymmetric case. This will in turn alter the waveforms in the final pre-explosion phase.
This has already been demonstrated by Kotake \etal \cite{Kotake09} in 3D parameterized studies.
3D simulations with all of the physics documented here are ongoing, and we look forward to reporting 
on their GW signatures in the near future.

{\bf Acknowledgements:} The authors would like to acknowledge the computational resources provided at the Leadership Computing Facility 
in the National Center for Computational Sciences at ORNL (INCITE Program) and at TACC (TG-MCA08X010). 
PM acknowledges partial support 
from NSF-PHYS-0855315, and PM and SWB acknowledge partial support from an NSF-OCI-0749204 award. AM, OEBM, 
PM, SWB, and WRH acknowledge partial support from a NASA ATFP award (07-ATFP07-0011). AM and WRH acknowledge 
support from the Office of Nuclear Physics, U.S. Department of Energy, and AM and OEBM acknowledge support from the Office 
of Advanced Scientific Computing Research, U.S. Department of Energy.

\end{document}